\pdfoutput=1
\documentclass[conference]{IEEEtran}

\usepackage{xspace}

\usepackage{esvect}
\usepackage{amssymb}
\usepackage{amsmath}
\usepackage{amsthm}
\newtheorem{definition}{Definition}

\newcommand{\cmscosdis}{CMS\textrm{-}cosSim\xspace}
\newcommand{\cbfdice}{CBF\textrm{-}Dice\xspace}
\newcommand{\cmsdice}{CMS\textrm{-}Dice\xspace}

\newcommand{\cbfrep}[1]{\ensuremath{#1^{\mathit{cbf}}}}
\newcommand{\cmsrep}[1]{\ensuremath{#1^{\mathit{cms}}}}

\usepackage{graphicx}

\ifCLASSINFOpdf
\else
\fi

\usepackage{amsmath}
\usepackage{xfrac}
\usepackage{url}

\begin{document}

\newcommand\copyrighttext{ 
	\Huge {IEEE Copyright Notice} \\ \\
	\large {Copyright (c) 2018 IEEE \\
		Personal use of this material is permitted. Permission from IEEE must be obtained for all other uses, in any current or future media, including reprinting/republishing this material for advertising or promotional purposes, creating new collective works, for resale or redistribution to servers or lists, or reuse of any copyrighted component of this work in other works.} \\ \\

	{\Large Published in: 2018 IEEE International Conference On Trust, Security And Privacy In Computing And Communications (IEEE TrustCom 2018), July 31 -- August 3, 2018} \\ \\
	DOI: 10.1109/TrustCom/BigDataSE.2018.00146 \\ \\
	\begin{small}
		Preprint from: \url{https://arxiv.org/abs/1805.07651}\\
		Print at: \url{https://doi.org/10.1109/TrustCom/BigDataSE.2018.00146}
	\end{small}
			
	\vspace{2cm}
	
	Cite as:\\
	\includegraphics[trim={2.075cm 23cm 9cm 3.1cm},clip]{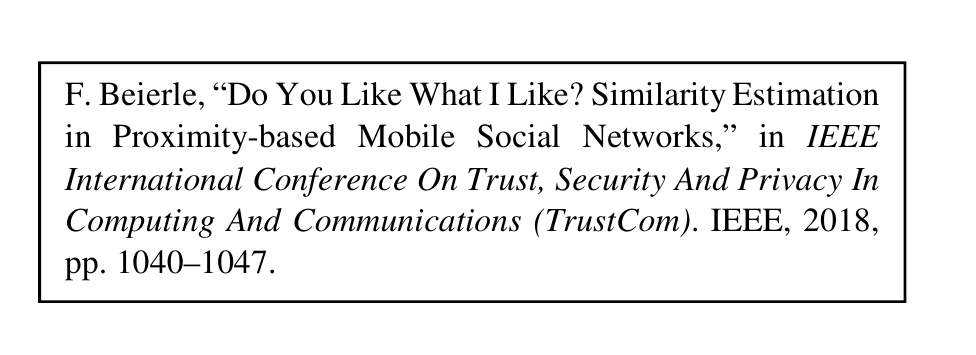}
	
	\vspace{1.5cm}
	
	BibTeX:\\ \\
	\includegraphics[trim={2.25cm 21cm 2cm 4cm},clip]{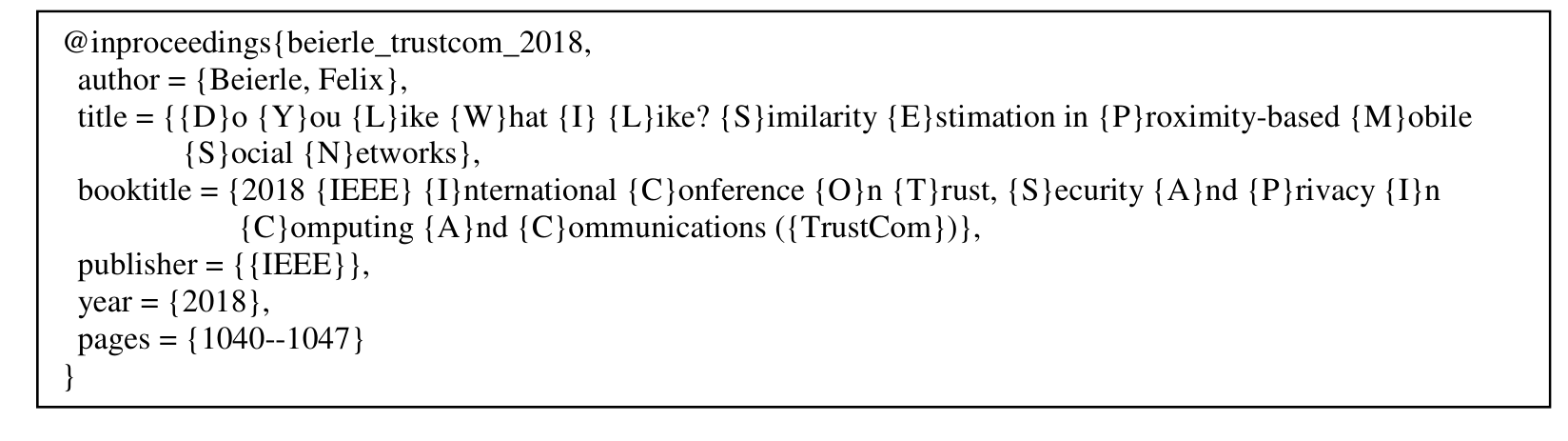}
}

\twocolumn[
\begin{@twocolumnfalse}
	\copyrighttext
\end{@twocolumnfalse}
]

\title{Do You Like What I Like? Similarity Estimation in Proximity-based Mobile Social Networks}

\author{
	\IEEEauthorblockN{Felix Beierle}
	\IEEEauthorblockA{Service-centric Networking\\
		Telekom Innovation Laboratories\\
		Technische Universit\"at Berlin\\
		Berlin, Germany\\
		beierle@tu-berlin.de}}

\maketitle

\begin{abstract}

While existing social networking services
tend to connect people who know each other,
people show a desire to also connect to yet unknown people in physical proximity.
Existing research shows that people tend to connect to similar people.
Utilizing technology in order to stimulate human interaction between
strangers,
we consider the scenario of two strangers meeting.
On the example of similarity in musical taste,
we develop a solution for the problem
of similarity estimation in proximity-based mobile social networks.
We show that a single exchange of a probabilistic data structure 
between two devices
can closely estimate the similarity of two users --
without the need to contact a third-party server.
We introduce metrics for fast and space-efficient approximation
of the Dice coefficient of two multisets -- based on the comparison of
two Counting Bloom Filters or two Count-Min Sketches.
Our analysis shows that utilizing a single hash function
minimizes the error when comparing these probabilistic data structures.
The size that should be chosen for the data structure depends on the
expected average number of unique input elements.
Using real user data, we show that a
Counting Bloom Filter with a single hash function and a length of 128 is sufficient
to accurately estimate the similarity between two multisets
representing the musical tastes of two users.
Our approach is generalizable for any other similarity estimation
of frequencies represented as multisets.

\end{abstract}


\IEEEpeerreviewmaketitle

\section{Introduction}
\label{sec:introduction}

In January of 2018,
the prime minister of the UK
appointed one of her ministers to focus on
issues related to loneliness\footnote{\url{https://www.theguardian.com/society/2018/jan/16/may-appoints-minister-tackle-loneliness-issues-raised-jo-cox}}.
This acknowledges the basic human need to form connections with
other people.
From a technological point of view, Online Social Networks (OSNs)
are typically used to reflect real-world social connections
and establish new ones.
Besides established global OSNs,
there is a rising market for services that explicitly focus on the connections
between people in proximity, i.e., neighborhood networks
like Nextdoor\footnote{\url{http://www.nextdoor.com/}} and its competitors.
We note the apparent desire and need for people
to form connections with those in physical proximity.
How can technology assist this need?

First, we observe that people use their smartphone to access social networking
services.
OSNs have shifted to Mobile Social Networks (MSNs).
Facebook, for instance, lists 1.15 billion mobile daily active users on average for 2016\footnote{\url{https://investor.fb.com/investor-news/press-release-details/2017/Facebook-Reports-Fourth-Quarter-and-Full-Year-2016-Results/default.aspx}}.
Among the most frequent concerns with established OSNs or MSNs
is the potential misuse of data and loss of user privacy.
This concern is heightened by the highly sensitive
user and sensor data that modern smartphones provide
(cf.\ \cite{beierlemobiSPC2018, beierlemobilesoft2018}).
One of the key reasons for these concerns is the centralized
architecture of the systems -- the social networking
service provider has all the data
and can use it
beyond the level to which users intended to share it with the service provider
\cite{falch_business_2009}.
By utilizing short range wireless interfaces, the
need for centralized servers in social networking
scenarios can be reduced.
Utilizing Bluetooth, WiFi Direct, or NFC,
smartphones can communicate directly with each other.

Consider the stimulation of connections between people who do not know each other yet,
utilizing a proximity-based MSN that takes into account
the described privacy concerns
of centralized social networking architectures.
Previous work points to what approach such a system can make in order
to foster user interaction:
psychological studies point out that any social network
is structured by homophily, which means
that people who are similar to each other tend to connect with
each other \cite{BGGS2017}.
How can we determine the similarity of two users with
mobile devices without contacting a central server?

When answering that question,
we have to consider the
applicability of the solution in a quickly changing mobile device-to-device
context, as well as implications imposed by
short-range technologies like NFC.
This implies the use of only small amounts of data and a limited number of necessary data exchanges.
A lot of the information that is relevant for the social profile
of a user is already available on the smartphone itself \cite{beierle_towards_2015}.
As the needed data is present and connectivity between devices is also given,
we focus on the question of how to process and compare the data
under the constraints of
bandwidth and computing limitations of mobile devices.

Based on the use case of two strangers meeting,
we develop a method of similarity estimation
for proximity-based MSNs.
Two users can quickly approximate their similarity
when meeting, without exchanging clear text data nor contacting any
central server.
While our approach is applicable to any other data that can be represented as a multiset,
in this paper,
we focus on one of the most typical features
of social profiles, the musical taste of the user.
Not only is listening to music one of the most typical
usages for smartphones \cite{smartphoneusage};
musical taste is, after gender, the most commonly disclosed
profile feature in Facebook \cite{farahbakhsh_analysis_2013}.
Musical taste is a common feature
that users tend to identify with, that thus can serve
as an appropriate feature
for similarity estimation or
can be used in the recommendation of new contacts.

In this paper, we present our approach
that allows the estimation of the similarity
of two users' musical tastes based on probabilistic data structures.
While approaches for set similarity estimation exist for the Bloom Filter (BF),
we develop an approach for Counting Bloom Filters (CBFs)
and Count-Min Sketches (CMSs), suitable for multisets.
We discuss our approach based on experiments done
with synthetic data and real user music listening history data.
We conclude with a concrete approach that is applicable for
multiset similarity estimations in device-to-device scenarios.

Hence, the main contributions of this paper are:
\begin{itemize}
	\item An approach to similarity estimations in proximity-based MSNs.
	\item The introduction of new comparison metrics for CBFs and for CMSs.
	\item An evaluation of the introduced metrics based on both synthetic and real data sets -- showing support for our approach to space-efficient similarity estimations.
\end{itemize}

\section{Related Work}
\label{sec:related-work}

Before the advent of smartphones,
a similar idea of social networking was described in
\cite{beach_whozthat?_2008}.
Here, users exchange identifiers of existing OSNs
with each other via Bluetooth and can look up
each other's information on the OSN.
Having the data already available on the smartphone
gives us the possibility to
directly compare data instead of relying on existing OSNs.
Some other papers present similar ideas,
utilizing Bluetooth and central servers
\cite{eagle_social_2005}
or
manually entered
interests to find user similarities
\cite {pietilainen_mobiclique:_2009}.
\emph{E-Smalltalker}
also follows the idea to share data via Bluetooth and describes a
so-called \emph{Iterative Bloom Filter} for finding the intersection
of two sets of topics of interest
\cite{yang_e-smalltalker:_2010}.
More recent work deals with
proximity-based mobile social networking:
with \emph{E-Shadow}, the user can see profiles of other users in proximity and
has to evaluate manually if he/she is interested in another user
without support for automatic similarity estimation
\cite{teng_e-shadow:_2014}.
In \emph{SANE},
the devices of users with similar interests are used for forwarding messages
\cite{mei_social-aware_2015},
while there is no system to stimulate interactions between users or offer recommendations
for new contacts.

Papers that focus more on the algorithmic side of this topic
often deal with the research areas of \emph{private set intersection}
or \emph{secure multi-party computation}.
Here, the application scenarios usually require a much higher
level of privacy than estimating similarity in the proximity-based
social networking scenario.
Especially the factor proximity reduces
potential attacker vectors.
Often, multiple data exchanges are necessary for
handshakes, key-exchanges, etc.\ \cite{dong_when_2013}.
Furthermore, some approaches rely on third parties
to perform homomorphic encryption
\cite{kerschbaum_outsourced_2012}
or need other peers to perform computations
\cite{tillmanns_privately_2015}.
While these third parties usually do not
learn anything about the two users,
they are not needed in our approach,
with which it is possible
to estimate the similarity of two users
with one single device-to-device data exchange.

\section{Background}
\label{sec:background}

\textbf{Multisets.}
A \emph{multiset} is a generalization of a set, allowing for multiple instances of each of its elements.
A series of events for which the
frequency is important -- but the order is not --
can be described as a multiset.
Take for example the visited locations of a user.
Every location or area can be described by a unique string.
Each visit adds one element of the corresponding string to the multiset.
Users with similar movement patterns will generate similar multisets.
In this paper, we focus on the musical taste of users.
Without the need to have the user explicitly enter this information,
we can just collect data about the songs a user listened to
\cite{beierle_privacy-aware_2016}.
Storing a unique string representation for each song for each time it is played
yields a multiset that represents the musical taste of the user.
In order to reduce the amount of data that needs to be exchanged,
we want to avoid sending clear text music playlists between clients.

\begin{figure}[h]
	\centering
	\includegraphics[width=0.5\columnwidth, trim = 3mm 4mm 0mm 4mm, clip=true]{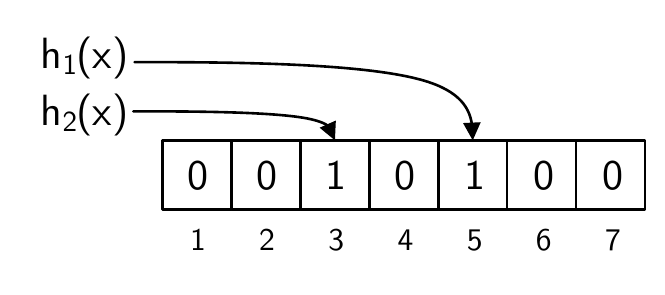}
	\caption{Visualization of a Bloom filter with a length of $n = 7$ and two hash functions ($k = 2$).}
	\label{fig:bf}
\end{figure}

\textbf{Bloom Filter (BF).}
Probabilistic data structures are able to represent large amounts of data space-efficiently.
Querying the data yields results with a certain probability.
The trade-off is between used memory and precision.
A BF yields probabilistic set membership \cite{bloom_space/time_1970-1}.
It consists of a bit vector with $n$ bits, initialized with zeros,
and $k$ pairwise independent hash functions,
each of which yields one position in the bit vector when hashing one element.
When adding an element to the set, all hash functions are applied,
yielding $k$ positions in the bit array, which then are switched to $1$.
We visualize a BF in Figure \ref{fig:bf}.
When querying for set membership,
the BF can answer whether the element is definitely not in the set
-- when the query element hashes to at least one $0$ in the bit vector --
or is likely in the set -- when all hash positions in the bit vector
are $1$. In the latter case the queried element is either in the set
or there is a collision with another element.

\begin{figure}[h]
	\centering
	\includegraphics[width=0.5\columnwidth, trim = 3mm 4mm 0mm 4mm, clip=true]{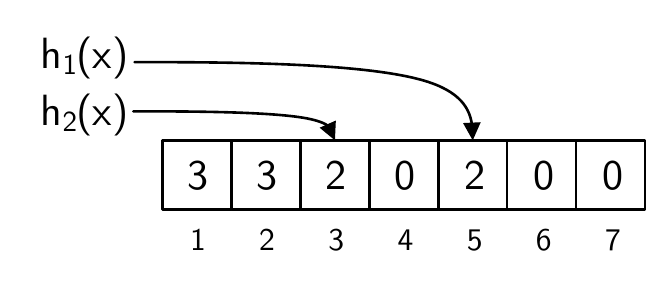}
	\caption{Visualization of a Counting Bloom filter with a length of $n = 7$ and two hash functions ($k = 2$).}
	\label{fig:cbf}
\end{figure}

\textbf{Counting Bloom Filter (CBF).}
An extension to the BF to adapt it to work with multisets
is to make each field in the bit vector not binary but a counter
\cite{fan_summary_2000}.
An example is shown in Figure \ref{fig:cbf}.
The resulting CBF can yield
a close estimation
of the cardinality of the queried element in a multiset.
Analogously to the false positives due to collisions in the original BF,
the estimated cardinality from the CBF represents an upper bound of an element's
cardinality in the original multiset.
In order to achieve a closer approximation of the cardinality of an element,
using multiple hash functions can be useful.
A single hash function can have collisions
and there can be collisions between hash functions.

In this paper, we use \emph{collision} instead of \emph{hash collision},
because the collisions that occur do not have to be hash collisions:
as each element has to be mapped to the length of the (C)BF
and not the entire namespace of the hash function,
there may be collisions even if there is no hash collision.
For example, two different hashes from hash function $h_1$
could be mapped to the same position of the (C)BF,
resulting in a collision but strictly speaking without having a \emph{hash} collision.
Utilizing multiple hash functions can help
reduce the impact of collisions: when
querying a CBF for the cardinality of an item $e$,
it is hashed with each hash function and
the lowest counter is returned.
Through the described collisions,
the yielded number is equal or higher than the true amount.

\begin{figure}[h]
	\centering
	\includegraphics[width=0.5\columnwidth, trim = 2mm 4mm 0mm 2mm, clip=true]{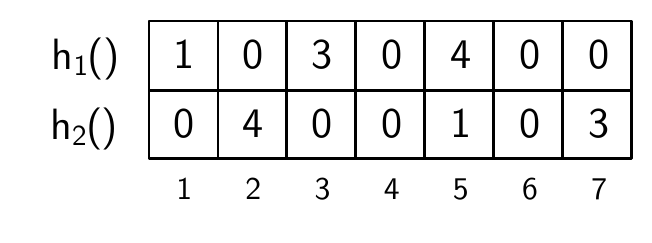}
	\caption{Visualization of a Count-Min Sketch with a width of $w = 7$ and depth of $d = 2$.}
	\label{fig:cms}
\end{figure}

\textbf{Count-Min Sketch (CMS).}
Another probabilistic data structure that works with multisets
is the Count-Min Sketch
\cite{cormode_improved_2004}.
It is often used to provide information about the frequency
of events in streams of data.
A CMS consists of $w$ columns and $d$ rows (cf.\ Figure \ref{fig:cms}).
Each field is initialized with $0$.
Each row is associated with one hash function.
Adding an element increments the counter at the
positions indicated by the hash functions.

It is worth stating the relationship between CBF and CMS:
adding the rows of a CMS yields a CBF.
A CBF with length $n$ and one single ($k = 1$) hash function $h$
is equal to a CMS with width $n = w$ and depth $1$,
given that the same hash function $h$ is used in the CMS:
\begin{equation}
\label{eq:cbf-cms}
{\mathit{CBF}_{n,k}} = {\mathit{CMS}_{w,d}}\quad |\quad n = w \quad\land\quad k=d=1
\end{equation}

\textbf{Comparison of two BFs.}
There are some papers dealing with the comparison
of two sets utilizing BFs
\cite{jain_using_2005}
\cite{schnell_privacy-preserving_2009}
\cite{donnet_path_2012}
\cite{alaggan_blip:_2012}.
In \cite{jain_using_2005}, the authors
compare two BFs with
a bit-wise $\mathit{AND}$.
A large number of $1$s in the result
indicates similarity.
The authors of \cite{schnell_privacy-preserving_2009}
calculate string similarity utilizing BFs
by creating n-grams, adding those into BFs
and using the Dice coefficient (see Equation \ref{eq:dice}).
The idea is that identical n-grams will hash to identical
positions in the bit array
as long as the same length and same hash functions are used.
In \cite{donnet_path_2012}, the authors
utilize BFs to estimate path similarity
for paths in computer networks.
They define a \emph{Bloom Distance},
which is the
logical $\mathit{AND}$ of both BFs, followed by counting
the number of $1$s in the results and dividing by the length
of a BF.
The authors of \cite{alaggan_blip:_2012}
use cosine similarity on two BFs to
determine similarity.
In \cite{yang_e-smalltalker:_2010},
the authors iteratively compare
two BFs.
They start with a small BF with a high false positive rate
and increase its size in a second round of comparisons
if the similarity value of the first round was above a pre-defined threshold.

\section{Metrics for Comparing CBFs and CMSs}
\label{sec:comparison}

Because in our scenario, we are dealing with multisets,
utilizing a BF and thus
leaving out the cardinality
of each element in the multiset would
not reflect the musical taste of the user anymore.
It is the cardinality of the element ("play count")
that indicates the number of times a specific song was played back.

In order to compare multisets, one can apply similar metrics like
suggested for the BF, i.e., cosine similarity and Dice coefficient.
Both cosine similarity (in case of positive values) and Dice coefficient
yield a value between $0$ and $1$, where $0$ indicates
no similarity and $1$ indicates sameness.
The cosine similarity for two vectors $v_1$ and $v_2$ is defined as:
\begin{equation}
\label{eq:cosine}
\mathit{cosSim}(v_1, v_2) = \frac{v_1 \cdot v_2}{||v_1|| \cdot  ||v_2||}
\end{equation}
The numerator indicates the dot product of $v_1$ and $v_2$.
The denominator indicates the multiplication of the
lengths of the two vectors: $||x|| = \sqrt{x_1^2+...+x_n^2}$.
Given a multiset $X$, we can interpret
the cardinalities of the elements in the multiset as elements of a vector.
Thus, for two multisets $X, Y$, we will just write $\mathit{cosSim}(X, Y)$, assuming an
appropriate vector representation of the cardinalities of the elements in $X$ and~$Y$.

The \emph{Dice coefficient} of two sets $A$ and $B$ is given by:
\begin{equation}
\label{eq:dice}
\mathit{Dice}(A, B) = \frac{2 \cdot |A \cap B|}{|A| + |B|}
\end{equation}
In order to obtain the Dice coefficient for two multisets $X$ and $Y$,
we can also use Equation \ref{eq:dice} by employing the cardinality 
and the intersection of multisets. The cardinality gives 
the sum of all occurrences of all elements in the multiset.
The intersection of two multisets $X$ and $Y$ can be determined by the
minimum function applied for each element $i$:
If $i \in^n X$ (denoting that $X$ has exactly $n$ instances of $i$) 
and $i \in^m Y$,
then the following holds for each element:
\begin{equation}
i \in^{min(n,m)} (X \cap Y)
\end{equation}

To the best of our knowledge, there is no research
specifically about the comparison of two CBFs or two CMSs.
A prerequisite for the pairwise comparison
is that the two data structures to be compared are
of the same length and use the same hash functions;
i.e.,
the same elements will always hash to the same positions
in the data structure.
Based on these prerequisites, we will now transform the idea of both cosine 
similarity and Dice coefficient to CBFs as well as to CMSs, yielding metrices 
for these data structures.

\subsection{Cosine Similarities for CBFs and CMSs}

As the data structure of a CBF is a vector, we can immediately use
Equation~\ref{eq:cosine} for defining the cosine similarity
$\mathit{cosSim}(P, Q)$ of two CBFs $P, Q$.

For the cosine similarity of CMSs, we view each CMS as a collection
of $d$ vectors (one vector each row), cf.\ Figure \ref{fig:cms}, and 
propose the following.\footnote{If $P$ (resp. $R$) is a CBF (resp. CMS), its positions will be denoted by $p_i$ (resp. $r_{ij}$).}

\begin{definition}[\cmscosdis]
\label{def:cmscosdis}
Let $R$ and $S$ be two CMSs with the same dimensions
$d \times w$ and utilizing the same hash functions.
The \emph{CMS cosine similarity} of $R$ and $S$ is given by:
\begin{align}
\mathit{\cmscosdis}(R,S) =
\frac{1}{d}  \cdot \sum_{i=1}^{d} \mathit{cosSim}(\vv{r_i},\vv{s_i})
\end{align}
where 
$\vv{r_i} = (r_{i1},\ldots,r_{iw})$
and 
$\vv{s_i} = (s_{i1},\ldots,s_{iw})$.
\end{definition}
Thus, the CMS cosine similarity of two CMSs averages the cosine similarity
of all corresponding rows. 

\subsection{Dice Coefficents for CBFs and CMSs}

The bitwise operations suggested for the comparisons of two BFs
do not work with CBF and CMS
as we have counters -- instead of binary values -- at each position, cf.\ Figure \ref{fig:cbf} and \ref{fig:cms}.
Therefore, we transfer the idea of the Dice coefficient for multisets to
CBFs as follows:

\begin{definition}[\cbfdice coefficient]
\label{def:cbfdice}
Let $P$ and $Q$ be two CBFs with length $n$ and utilizing the same 
hash functions.
The \emph{CBF dice coefficient} of $P$ and $Q$ is given by:
\begin{equation}
\label{eq:cbfSim}
\mathit{\cbfdice}(P,Q) =
\frac{ 2 \cdot \sum_{i=1}^{n}  min(p_{i}, q_{i})} {\sum_{i=1}^{n} p_{i} + q_{i}}
\end{equation}
\end{definition}

Note that 
in order to approximate the numerator for the multiset Dice coefficient,
the CBF dice coefficient in Equation~\ref{eq:cbfSim}
applies the minimum function for each position, and
for the denominator, the cross sum of both CBFs is used.

Extending the Dice coefficient to CMSs reuses the dice coefficient for CBFs.

\begin{definition}[\cmsdice coefficient]
\label{def:cmsdice}
Let $R$ and $S$ be two CMSs with the same dimensions
$d \times w$ and utilizing the same hash functions.
The \emph{CMS dice coefficient} of $R$ and $S$ is given by:
\begin{equation}
\label{eq:cmsSim}
\mathit{\cmsdice}(R, S)
= \frac{1}{d} \cdot \sum_{i=1}^{d}
\frac{ 2 \cdot \sum_{j=1}^{w} min(r_{ij}, s_{ij})} {\sum_{j=1}^{w} r_{ij} + s_{ij}}
\end{equation}
\end{definition}

Hence, the Dice coefficient of two CMSs utilizes the average of the CBF Dice
coefficient of each pair of corresponding rows.

\subsection{Comparing Mulisets via CBFs and CMSs}

Given two multisets $X$ and $Y$, we can now estimate their similarity via
CBFs or via CMSs.
Let $\cbfrep{X}$ and $\cbfrep{Y}$ be CBFs 
(of the same length and using the same hash functions)
and let 
$\cmsrep{X}$ and $\cmsrep{Y}$ be CMSs 
(of the same dimensions and using the same hash functions)
for $X$ and $Y$.
In the following evaluation, we show how we can use
\begin{equation}
\label{approx:cbfdice}
\mathit{\cbfdice}(\cbfrep{X},\cbfrep{Y})
\end{equation}
and
\begin{equation}
\label{approx:cmsdice}
\mathit{\cmsdice}(\cmsrep{X},\cmsrep{Y})
\end{equation}
to approximate
\begin{equation}
\mathit{Dice}(X,Y).
\end{equation}
Likewise, we investigate the approximation of 
\begin{equation}
\mathit{cosSim}(X,Y)
\end{equation}
by
\begin{equation}
\label{approx:cbfcosdis}
\mathit{cosSim}(\cbfrep{X},\cbfrep{Y})
\end{equation}
and
\begin{equation}
\label{approx:cmscosdis}
\mathit{\cmscosdis}(\cmsrep{X},\cmsrep{Y}).
\end{equation}

Using any of the approximations given in 
\eqref{approx:cbfdice},
\eqref{approx:cmsdice},
\eqref{approx:cbfcosdis}, or
\eqref{approx:cmscosdis}
will typically yield a fast and space efficient comparison   
of the multisets $X$ and $Y$. Our evaluation on both synthetic and on
real data shows that good approximations may already be achieved when
using small CBFs or CMSs.

\section{Experimental Results}
\label{sec:experiment}

For the evaluation, we use both a synthetic data set and real user music listening histories.
The synthetic data set
\begin{equation*}
\label{eq:tmp}
\mathit{SD} = { A_{r}, A_{0}, ..., A_{1000}}
\end{equation*}
consists of $1{,}002$ multisets of strings.
For the strings contained in each of those multisets, we use
random ASCII strings that are $10$ characters long.
As the strings are entered into the hash functions of CBF and CMS, we could
have picked any other random string to achieve the same effects.
Each multiset has $66.9$ unique entries on average.
Given a multiset of random strings $A_{r}$, the other multisets
$A_{i}$ are chosen such that comparing $A_{r}$ to the other multisets yields Dice
coefficients of $0.000$ to $1.000$ in increments of $0.001$, i.e.,
\begin{equation*}
\label{eq:tmp2}
Dice(A_{r}, A_{i}) = i * 0.001
\end{equation*}

For the real data RD, we used the taste profile
subset\footnote{\url{https://labrosa.ee.columbia.edu/millionsong/tasteprofile}}
of the million song data set \cite{bertin-mahieux_million_2011}.
In order to have appropriate data for comparisons, we chose a subset of active users
who each listened to at least $50$ distinct songs.
The subset contains $1{,}865$ distinct users,
$14{,}867$ distinct song titles, and $122{,}389$ recorded plays.
In order to be able to visualize the results,
we chose a subset of these users that yields a
range of different similarity values.
To enter data into the data structures, we
built a unique string for each song.
RD consists of roughly $4{,}000$ multiset comparisons.
On average, each multiset has $63.8$ unique entries.

\subsection{Synthetic Data SD / Comparing CMSs}

\begin{figure}[h]
	\centering
	\includegraphics[width=0.84\columnwidth, trim = 8mm 2mm 13mm 13mm, clip=true]{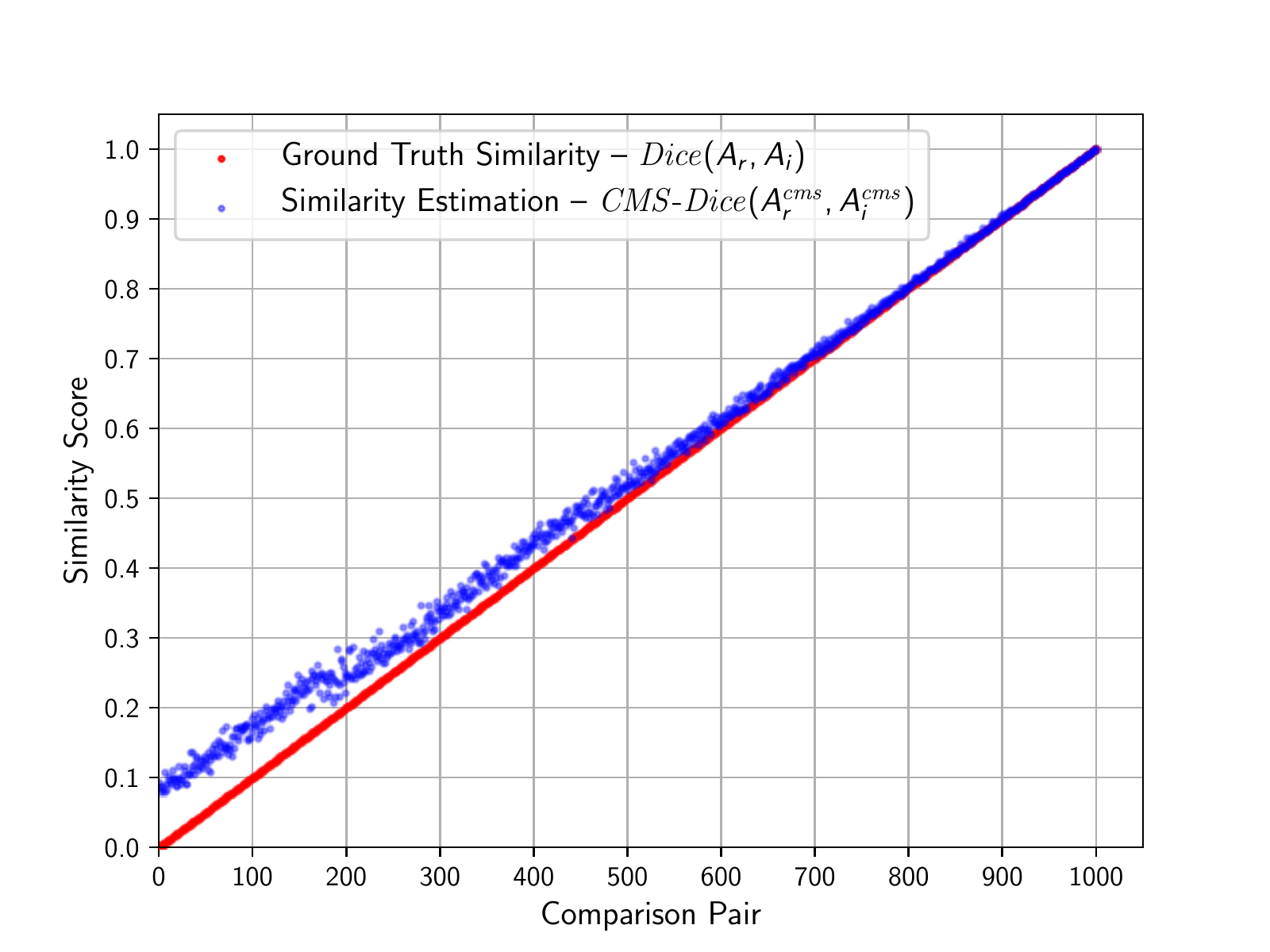}
	\caption{Similarity scores for ground truth $\mathit{Dice}(A_{r},A_{i})$ in red and estimation $\mathit{\cmsdice}(\cmsrep{A_{r}},\cmsrep{A_{i}})$ in blue using a CMS with $w=400$ columns and $d=10$ rows, using synthetic data set SD.}
	\label{fig:syn-cms-dice}
\end{figure}

For evaluating \cmsdice on SD, we start with a CMS encoding
of SD with $w = 400$ columns and $d = 10$ rows.
For each $A_{i} \in SD$,
let $\cmsrep{A_{i}}$ be the corresponding CMS for $A_{i}$.
Figure \ref{fig:syn-cms-dice} illustrates the comparison of the Dice coefficient
of the multisets as ground truth -- $\mathit{Dice}(A_{r},A_{i})$ -- with the \cmsdice of the CMS
representation -- $\mathit{\cmsdice}(\cmsrep{A_{r}},\cmsrep{A_{i}})$.
The Dice coefficient is plotted in red.
The x-axis indicates the multiset pair combination
that is compared -- sorted by Dice coefficient.
The y-axis gives the similarity score.
The blue dots represent the \cmsdice similarity score.

\begin{figure}[h]
	\centering
	\includegraphics[width=0.84\columnwidth, trim = 8mm 2mm 13mm 13mm, clip=true]{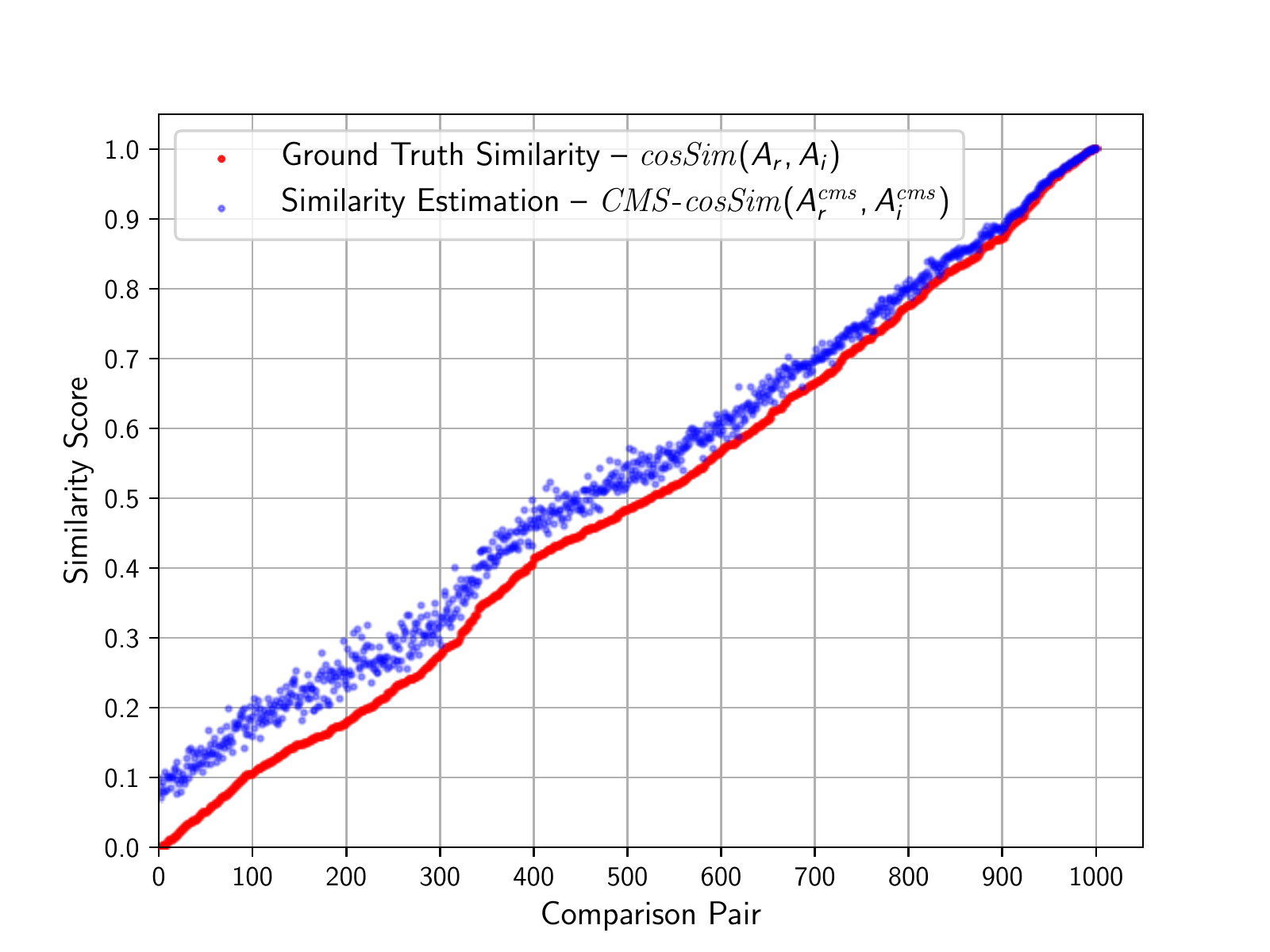}
	\caption{Similarity scores for ground truth $\mathit{cosDis}(A_{r},A_{i})$ in red and estimation $\mathit{\cmscosdis}(\cmsrep{A_{r}},\cmsrep{A_{i}})$ in blue using a CMS with $w=400$ columns and $d=10$ rows, using synthetic data set SD.}
	\label{fig:syn-cms-cosine}
\end{figure}

In Figure \ref{fig:syn-cms-cosine}, we give the same plot for cosine similarities.
We compare the cosine similarity of the multisets as
ground truth -- $\mathit{cosDis}(A_{r},A_{i})$ -- with the
cosine similarity of the CMS representation --
$\mathit{\cmscosdis}(\cmsrep{A_{r}},\cmsrep{A_{i}})$.
The first observation we make is that Dice and cosine
measurement yield almost identical results, which
can be seen by the red lines in Figure \ref{fig:syn-cms-dice} and \ref{fig:syn-cms-cosine}
having almost identical slopes.
Therefore, for the rest of this paper, we focus on the Dice coefficient.
Another observation we make is that the similarity estimation
by \cmsdice is always slightly higher than or equal to the Dice coefficient ground truth
-- so the similarity between two multisets is always correctly estimated or over-estimated
due to collisions,
never underestimated.

\begin{figure}[b]
	\centering
	\includegraphics[width=0.84\columnwidth, trim = 32mm 4mm 23mm 20mm, clip=true]{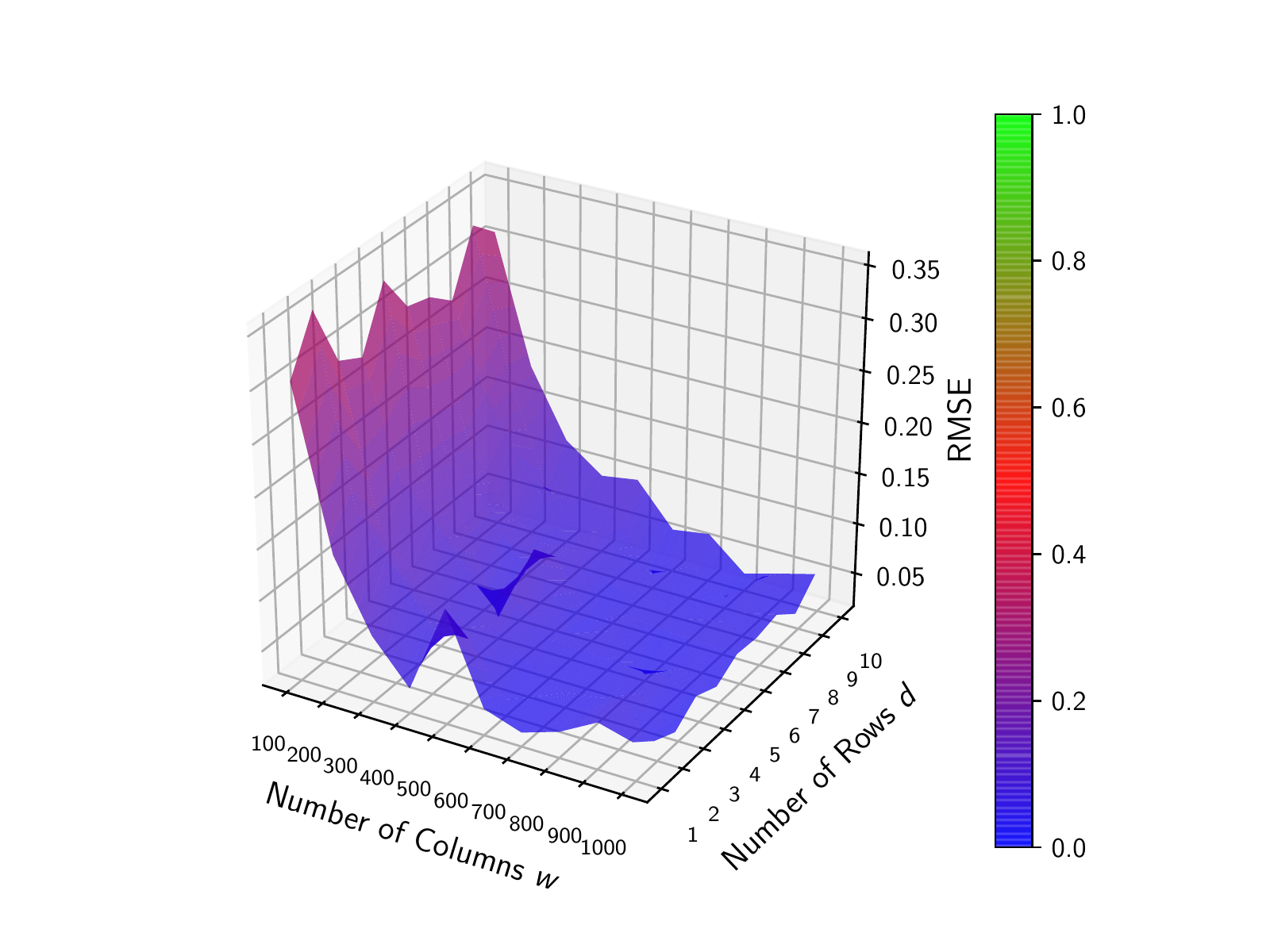}
	\caption{RMSEs of similarity estimation $\mathit{\cmsdice}$ for CMSs of different sizes, using synthetic data set SD and $\mathit{Dice}$ coefficient as ground truth.}
	\label{fig:syn-rmse-cms-dice}
\end{figure}

Typically, when using a CMS, the user wants to
perform queries and get accurate results.
The values for $w$ and $d$ are chosen accordingly.
In our case, we just want to perform a
similarity estimation
without querying for specific entries.
We investigate what influence the number of columns
$w$ and the number of rows $d$ have on the estimation of similarity.
In order to do so, we plot the same comparisons of the synthetic data
for different values of $w$ and $d$.
For each combination, we calculate the root mean square error (RMSE)
of the similarity estimation by \cmsdice from the Dice coefficient of
the ground truth.
The RMSE quantifies to what extent the similarity estimation
differs from the ground truth similarity score.
The lower the RMSE, the better the approximation of the Dice coefficient.
Based on $100{,}100$ comparisons, we calculate
the RMSE for different combinations of
$w$ (x-axis) and $d$ (y-axis) values.
The result is given in Figure \ref{fig:syn-rmse-cms-dice}.
With an increasing number of columns, the RMSE decreases.
The number of rows does not significantly influence the RMSE:
increasing the number of rows
does not reduce the RMSE
of the similarity estimation.

\subsection{Synthetic Data SD / Comparing CBFs}

\begin{figure}
	\centering
	\includegraphics[width=0.84\columnwidth, trim = 32mm 4mm 23mm 20mm, clip=true]{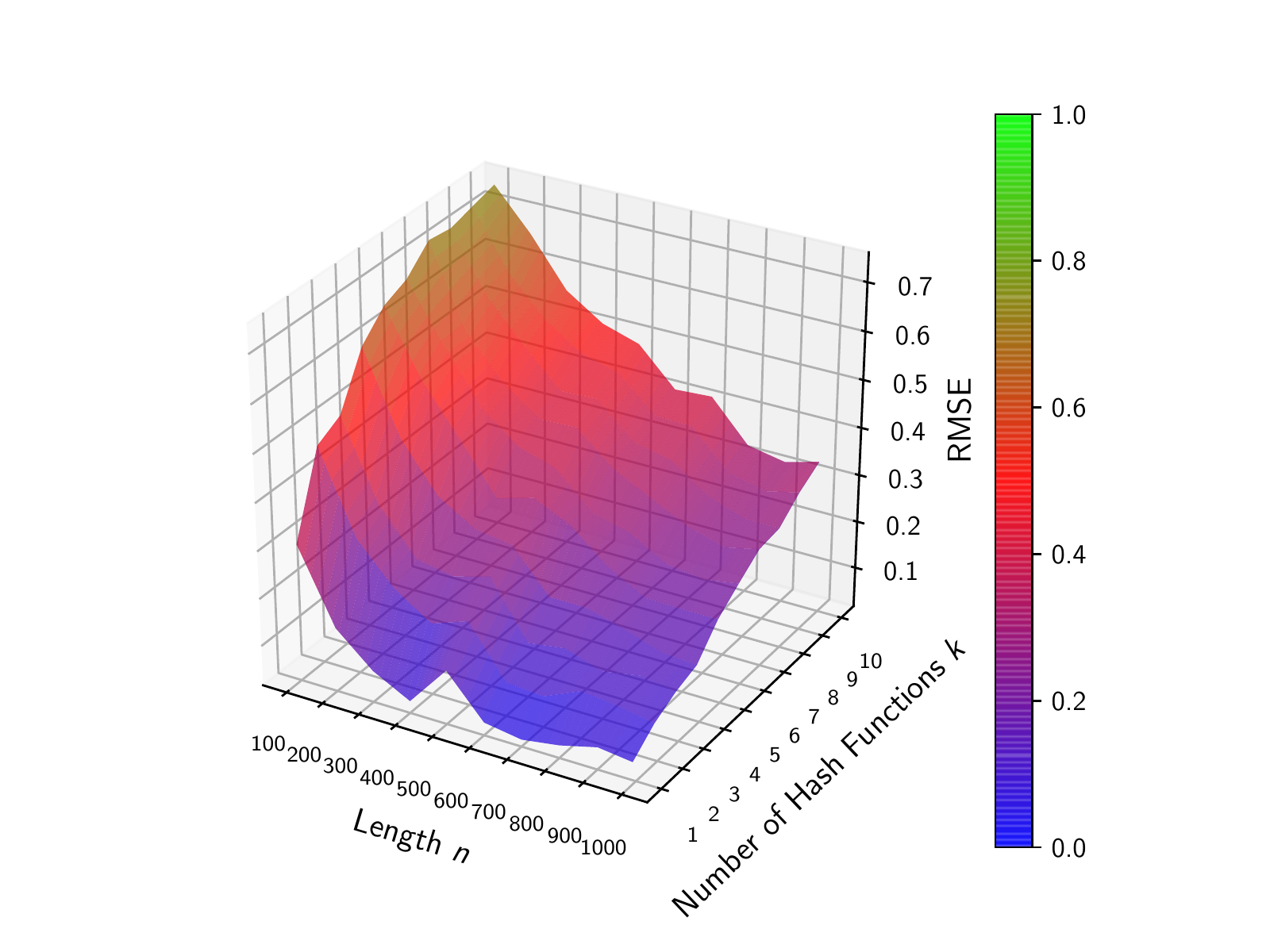}
	\caption{RMSEs of similarity estimation $\mathit{\cbfdice}$ for CBFs of different sizes, using synthetic data set SD and $\mathit{Dice}$ coefficient as ground truth.}
	\label{fig:syn-rmse-cbf-dice}
\end{figure}

Visualizing the RMSEs for similarity estimation by \cbfdice with different
length $n$ and number of hash functions $k$,
we get Figure \ref{fig:syn-rmse-cbf-dice}.
Increasing the length of the CBF reduces the average error of \cbfdice
while increasing the number of hash functions increases the error.

\subsection{Real Data RD}

\begin{figure}
	\centering
	\includegraphics[width=0.84\columnwidth, trim = 32mm 4mm 23mm 20mm, clip=true]{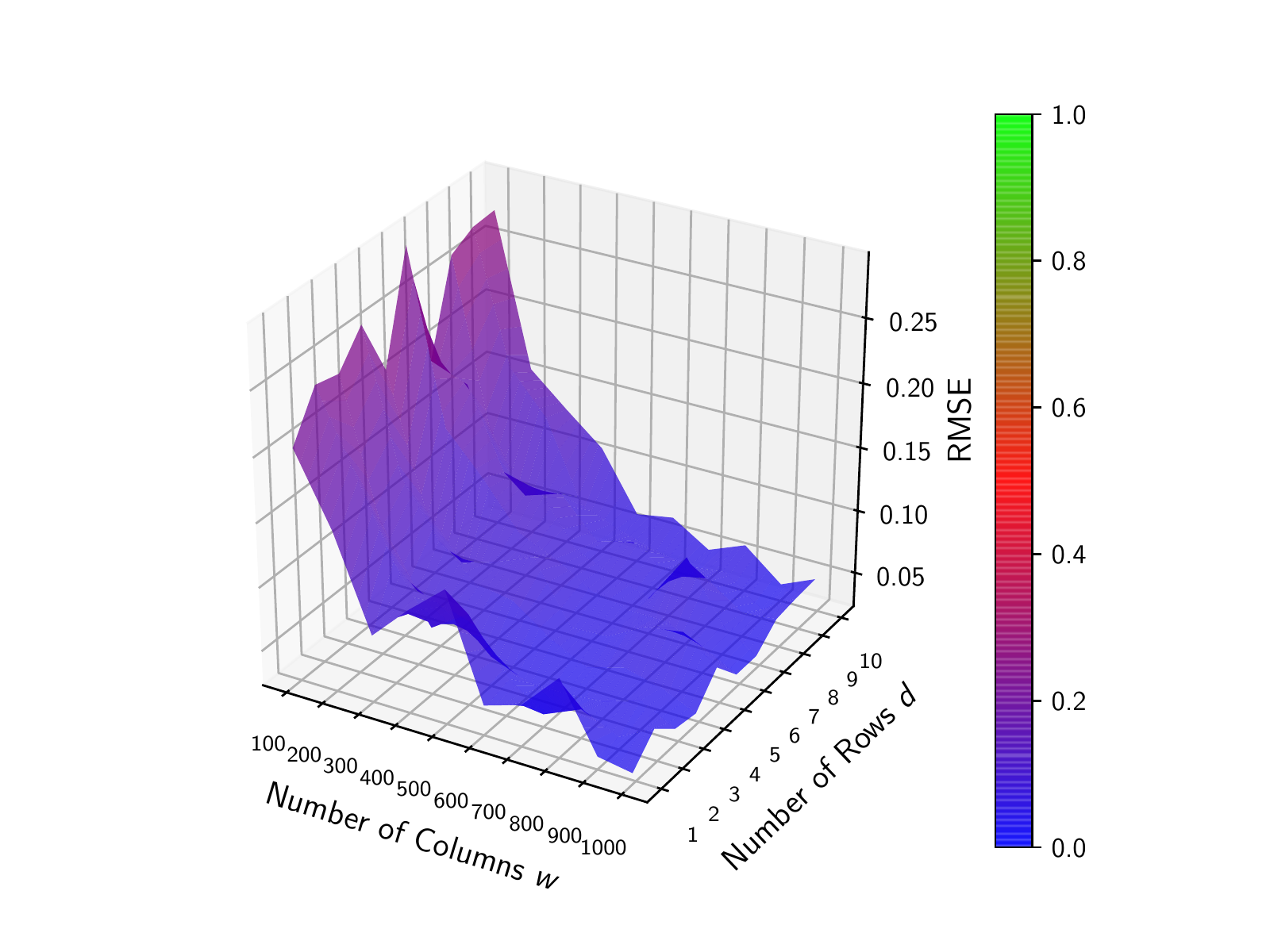}
	\caption{RMSEs of similarity estimation $\mathit{\cmsdice}$ for CMSs of different sizes, using real data set RD and $\mathit{Dice}$ coefficient as ground truth.}
	\label{fig:real-rmse-cms-dice}
\end{figure}

\begin{figure}
	\centering
	\includegraphics[width=0.84\columnwidth, trim = 32mm 4mm 23mm 20mm, clip=true]{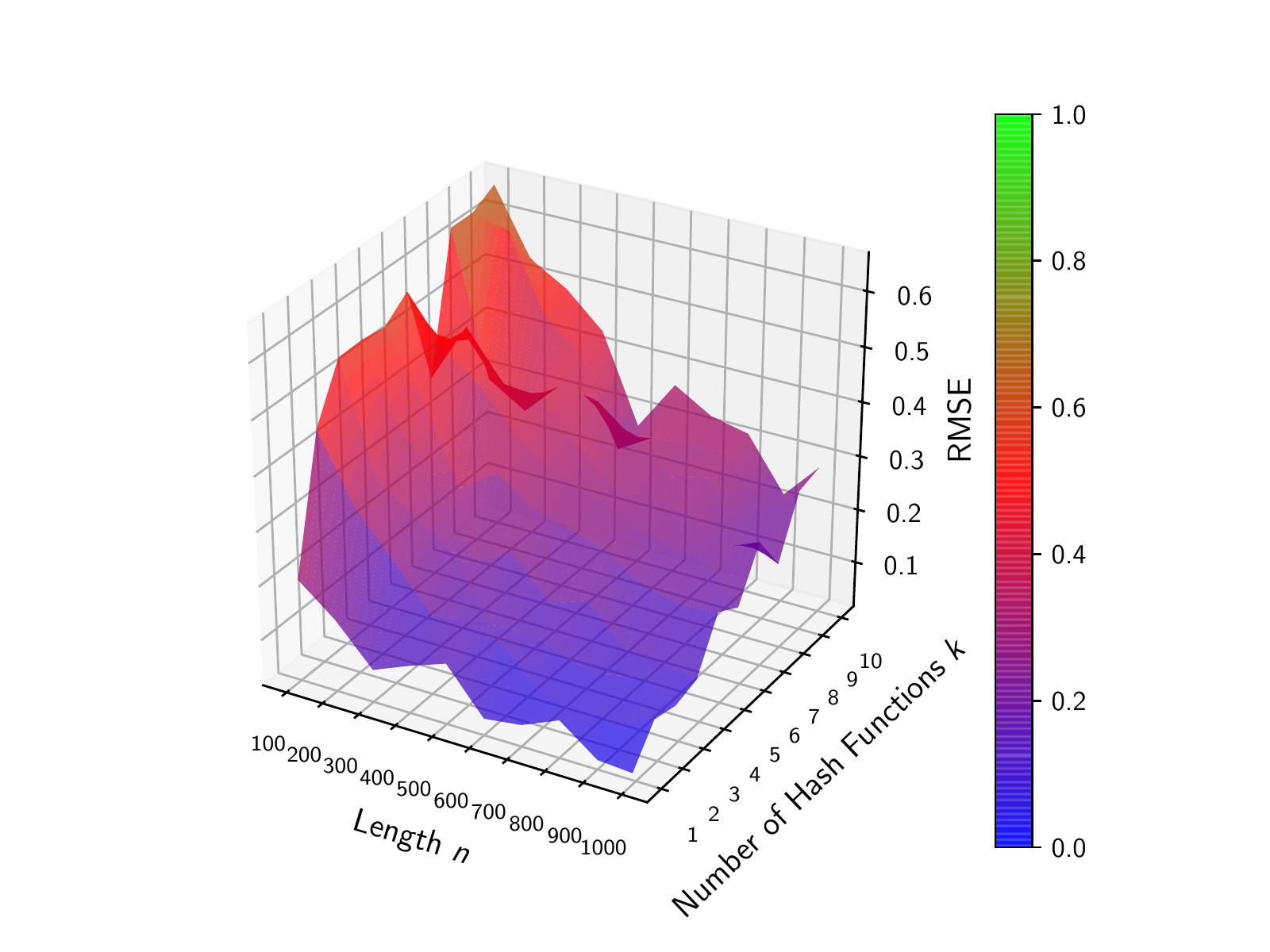}
	\caption{RMSEs of similarity estimation $\mathit{\cbfdice}$ for CBFs of different sizes, using real data set RD and $\mathit{Dice}$ coefficient as ground truth.}
	\label{fig:real-rmse-cbf-dice}
\end{figure}

\begin{figure}
	\centering
	\includegraphics[width=0.84\columnwidth, trim = 8mm 2mm 13mm 13mm, clip=true]{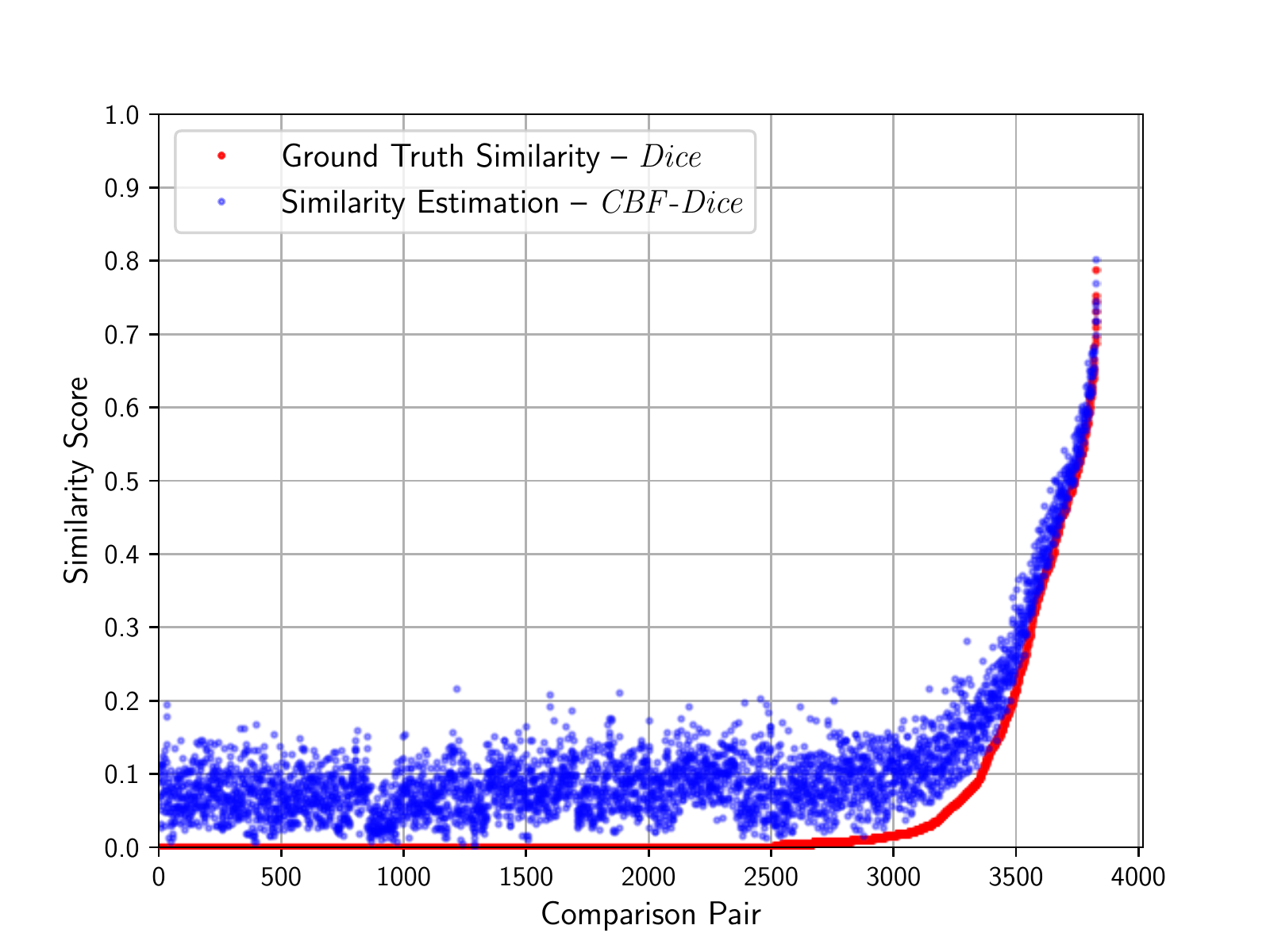}
	\caption{Similarity scores for ground truth $\mathit{Dice}$ in red and estimation $\mathit{\cbfdice}$ in blue using a CBF/CMS with length $400$ and $1$ hash function, using real data set RD.}
	\label{fig:real-cbf-n400-k1-dice}
\end{figure}

Using the real data set RD, our findings using synthetic data SD are confirmed.
Figure \ref{fig:real-rmse-cms-dice} shows the RMSEs
for \cmsdice, and Figure \ref{fig:real-rmse-cbf-dice} for \cbfdice.
We achieve the highest accuracy and simultaneously the lowest
memory size by using a CMS with one row, which is
a CBF with one hash function (cf.\ Equation \ref{eq:cbf-cms}).
In Figure \ref{fig:real-cbf-n400-k1-dice}, we visualize the estimation error with \cbfdice
(which is equal to \cmsdice in this case)
utilizing this data structure with a length of $400$.
We can see the error of each similarity estimation and
can see that we never underestimate the similarity.
Looking at the first $3{,}000$ comparison pairs,
the values for the ground truth similarity scores
are very low.
The values for the similarity estimation by \cbfdice range
roughly from $0$ to $0.2$.
For the remaining comparisons,
we observe that the higher the ground truth similarity score
is, the lower is the error range by the \cbfdice estimation.

\section{Discussion}
\label{sec:discussion}

In order to discuss the experimental results,
we start by analyzing the regular BF.
Consider a regular BF with one hash function, utilized for comparing sets.
If two sets are equal, the BFs should be equal and even the smallest length
yields the correct estimation.
Regarding memory size (length of the BF),
comparing two disjunct sets is the worst case scenario:
estimating the similarity of two disjunct sets by
performing $\mathit{AND}$ on the two BFs should yield $0$ for every position.
There is one factor that introduces a deviation from $0$:
the number of unique inputs for a given length of the BF.
Because of the limited length of the BF,
several elements are hashed to the same position in the bit vector,
even if there is no hash collision
(see description in Section \ref{sec:background}).
By increasing the length of the BF, the probability for such collisions
is reduced.
The more unique elements are entered into the BF,
the more positions are set to $1$.
Thus, the more unique elements are in at least one of the sets,
the longer the bit vector should be if a small error in
estimating the similarity is desired.

When comparing two disjunct sets with cardinalities
$g$ and $h$, the BFs have to have at least a length
of $g+h$ in order to theoretically be able to correctly
estimate a similarity of $0$.
Now imagine increasing the number of hash functions.
It increases the error:
more bits are set to $1$,
which creates a higher similarity estimation.
In the following, we show that these conclusions
are also true for the CMS and CBF.

As described in Section \ref{sec:background},
when the goal is to query a CBF or CMS for cardinality of an
element of a multiset, the user profits from utilizing
multiple hash functions.
In our scenario of similarity estimation,
we do not need to query for specific items
and do not profit from multiple hash functions in the same way.
As described above for the BF, the opposite is true for the CBF:
both Figure \ref{fig:syn-rmse-cbf-dice}
and \ref{fig:real-rmse-cbf-dice} indicate
the trend that the more hash functions we use,
the worse the error is.
This is because of the collisions:
multiple elements are mapped to the same positions
in the bit vector.
The more hash functions we use, the more collisions
there are.
This can be compensated by increasing the length
of the CBF, which would just
unnecessarily increase the needed memory size.

Regarding the CMS, we increase the number of hash functions
by increasing the number of rows (see Figures \ref{fig:syn-rmse-cms-dice}
and \ref{fig:real-rmse-cms-dice}).
However, we do not see an increase in error when using more
hash functions.
This is because each hash function corresponds to one row.
The probability of collisions is the same in each row
and the average of the row-wise similarities
calculated by \cmscosdis and \cmsdice contains this error.
Considering memory size and computation time,
we should use just one single row.

The best and worst case for similarity estimation are the same
as described for the BF:
the higher the real similarity, the lower is the RMSE.
The same elements definitely hash to the same positions.
Only those elements not present in the other multiset
introduce an error in the similarity estimation
through collisions.
Thus, the more dissimilar the multisets are,
the larger the potential error.
This can best be seen in Figure \ref{fig:real-cbf-n400-k1-dice}.
Note how the spread of blue dots (similarity estimations by \cbfdice)
spans a larger part of the y-axis (similarity score)
for lower ground truth similarity scores (red dots).
This means that for lower similarities,
there are higher errors.
All errors are produced by collisions
and lead to an overestimation of similarity.

Compared to the BF comparison, for the CBF and CMS, the cardinality of
each element is the additional factor to consider.
For the regular BF, each collision has the same effect on the error.
Using a CBF or CMS, the influence a collision has on the error
of the similarity estimation is based on how the
data of the multiset is distributed.
If two users listen to two different songs very frequently
and those two songs are mapped to the same position in the data structure,
then the error in the similarity estimation can be significant.
The error is less significant if the
collision occurs for two different songs the two users
listened to less frequently.

We conclude that 
a general approach for
estimating the similarity of two multisets in proximity-based mobile applications
can be:
\begin{itemize}
	\item use one-hash CBF / one-row CMS as a data structure
	\item estimate the average number of unique input elements
	\item define an appropriate threshold for the given scenario
\end{itemize}
We showed that the one-hash CBF gives the best estimation while having the smallest memory size.
Our discussion showed that the average number of unique input elements is the relevant factor for how well the estimation is.
Based on this number, we can pick the length of the CBF.
We showed that a length of twice the average unique inputs is necessary
to theoretically still be able to estimate with full accuracy
in the worst case scenario of disjunct multisets.
Lastly, after performing the similarity estimation, one should have a threshold
to be able to tell if the result should be regarded as significant.

\begin{figure}[h]
	\centering
	\includegraphics[width=0.84\columnwidth, trim = 8mm 2mm 13mm 13mm, clip=true]{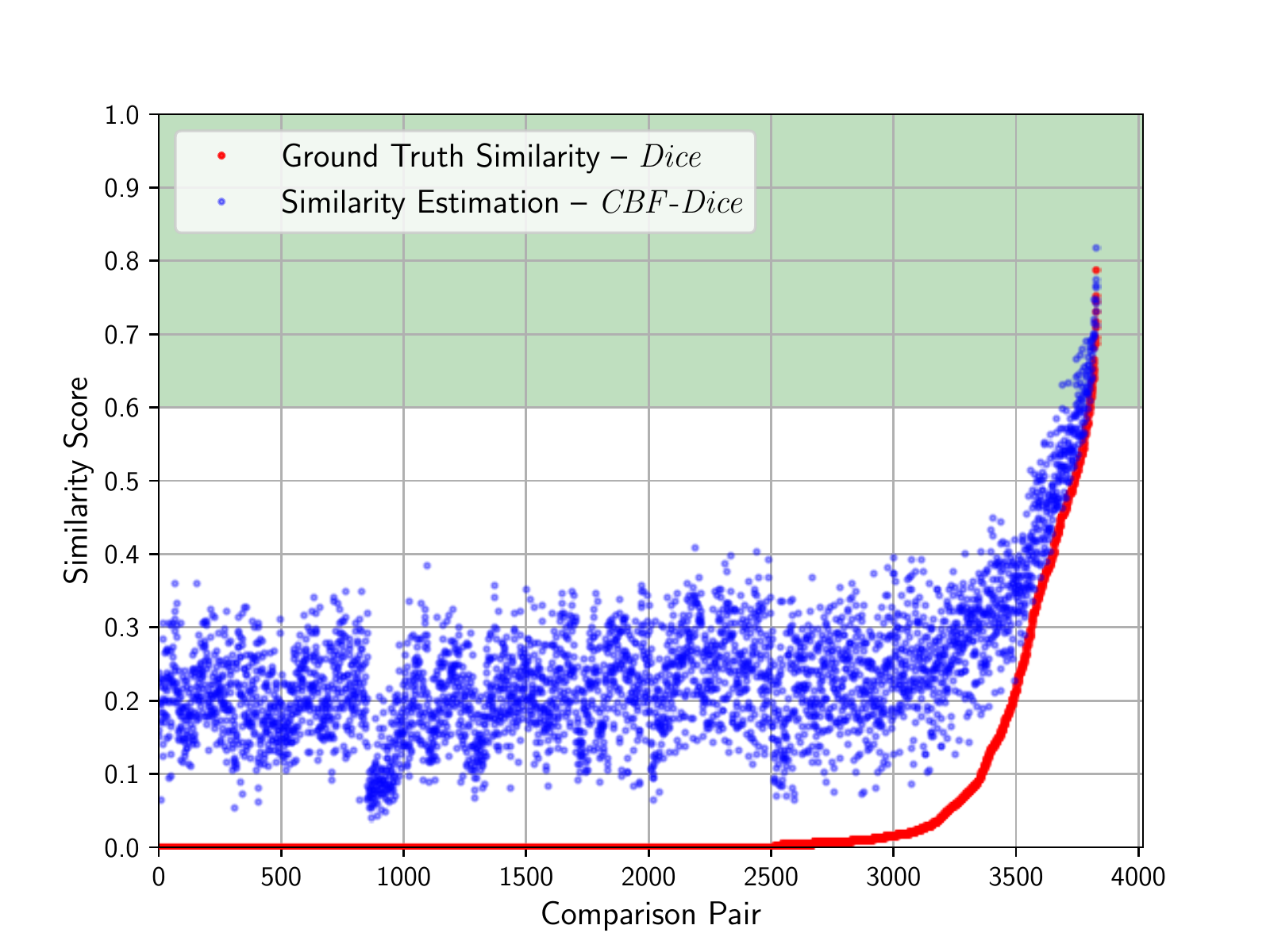}
	\caption{Similarity scores for ground truth $\mathit{Dice}$ (red) and estimation $\mathit{\cbfdice}$ (blue) using one-hash CBF (length $128$), using real data set RD. The green area indicates similarity scores above the desired threshold of $0.6$.}
	\label{fig:real-cbf-n128-k1-dice}
\end{figure}

Taking the music listening histories in our scenario,
we have an average unique input of $63.8$ elements
and regard similarities above $0.6$ to be relevant.
Picking a size of twice the average unique input
gives a length of the data structure of $128$.
We plot the ground truth (red) and the \cbfdice similarity estimation (blue)
in Figure \ref{fig:real-cbf-n128-k1-dice}.
The green area marks similarity scores above $0.6$.
We observe some false positives: the blue dots in the green area
that correspond to red dots below the green area.
The left-most blue dot in the green area indicates the largest error.
Here, we estimate a significant similarity of $\geq0.6$
while the ground truth value is about $0.5$.
If we need more accurate results, we choose a larger
length, for example like in Figure \ref{fig:real-cbf-n400-k1-dice}.
A positive side effect of the larger errors for low values is
that it somewhat provides privacy through lack-of-accuracy:
estimating a similarity of $0.4$ corresponds to
a ground truth value of between $0$ to about $0.4$ -- we cannot make an accurate
assumption about the actual similarity.

\section{Conclusion}
\label{sec:conclusion}

In this paper,
we approached the problem of multiset similarity estimation
in the scenario of proximity-based MSNs.
We developed the comparison
metrics \cbfdice and \cmsdice for
the similarity estimation of two CBFs and two CMSs.
Applying these metrics, we can approximate the Dice coefficient
when comparing two multisets.
We evaluated our approach with both
synthetic data and real music listening history data.

Our results show that
the more hash functions we utilize in a data structure,
the higher is the error in the estimation.
The larger the data structure is, the smaller is the error.
We achieve the lowest error when utilizing a
one-hash CBF / one-row CMS.
Here, we minimize the number of collisions
by using only one hash function.
The collisions are the source of the error in the similarity estimation.
In general, the higher the real similarity, the better is the estimation.

The data structure one-hash CBF is appropriate for the given scenario
of similarity estimation between two users,
requiring only a single data exchange
between two smartphones.
We described the general approach for assessing
the appropriate size of the data structure
by estimating the average number of unique input elements
as well as defining a threshold for the similarity score.
Using the real user music listening histories with a mean of about $64$
unique entries,
we showed that a one-hash CBF with length $128$ (twice the average unique inputs) suffices to
accurately estimate the similarity between two multisets.

While we presented the scenario of two strangers meeting
and quickly determining the similarity of their musical tastes,
our approach can be applied in a variety of other scenarios.
In general, any two systems that log any events
can be compared with our approach.
Utilizing \cbfdice, one can perform fast and space-efficient similarity estimations
of the two systems in terms of the frequencies of the logged events.

For future work in the proximity-based MSN scenario,
potential attack scenarios like malicious users should be addressed.
Furthermore, an implementation for mobile devices
can help evaluate the performance of our proposed approach.
For the scenario of stimulating interaction between strangers,
additional features besides musical taste should be considered,
e.g., visited locations.

\section*{Acknowledgment}
This work has received funding from project
DYNAMIC\footnote{\url{http://www.dynamic-project.de}} (grant No 01IS12056), which is funded
as part of the Software Campus initiative by the German
Federal Ministry of Education and Research (BMBF).
We are grateful for the support provided by
Niklas Lensing,
Bianca L\"uders,
Peter Ruppel,
Boris Lorbeer,
Sandro Rodriguez Garzon,
Martin Westerkamp,
Kai Grunert,
Tanja Deutsch,
Bernd Louis,
and Axel K\"upper.

\bibliographystyle{IEEEtran}

\begin{thebibliography}{10}
\providecommand{\url}[1]{#1}
\csname url@samestyle\endcsname
\providecommand{\newblock}{\relax}
\providecommand{\bibinfo}[2]{#2}
\providecommand{\BIBentrySTDinterwordspacing}{\spaceskip=0pt\relax}
\providecommand{\BIBentryALTinterwordstretchfactor}{4}
\providecommand{\BIBentryALTinterwordspacing}{\spaceskip=\fontdimen2\font plus
\BIBentryALTinterwordstretchfactor\fontdimen3\font minus
  \fontdimen4\font\relax}
\providecommand{\BIBforeignlanguage}[2]{{%
\expandafter\ifx\csname l@#1\endcsname\relax
\typeout{** WARNING: IEEEtran.bst: No hyphenation pattern has been}%
\typeout{** loaded for the language `#1'. Using the pattern for}%
\typeout{** the default language instead.}%
\else
\language=\csname l@#1\endcsname
\fi
#2}}
\providecommand{\BIBdecl}{\relax}
\BIBdecl

\bibitem{beierlemobiSPC2018}
F.~Beierle, V.~T. Tran, M.~Allemand, P.~Neff, W.~Schlee, T.~Probst, R.~Pryss,
  and J.~Zimmermann, ``{C}ontext {D}ata {C}ategories and {P}rivacy {M}odel for
  {M}obile {D}ata {C}ollection {A}pps,'' \emph{{P}rocedia {C}omputer
  {S}cience}, 2018 (to appear).

\bibitem{beierlemobilesoft2018}
------, ``{TYDR} -- {T}rack {Y}our {D}aily {R}outine. {A}ndroid {A}pp for
  {T}racking {S}martphone {S}ensor and {U}sage {D}ata,'' in \emph{2018
  {ACM/IEEE} 5th {I}nternational {C}onference on {M}obile {S}oftware
  {E}ngineering and {S}ystems {(MOBILESoft '18)}}.\hskip 1em plus 0.5em minus
  0.4em\relax ACM, 2018, pp. 72--75.

\bibitem{falch_business_2009}
M.~Falch, A.~Henten, R.~Tadayoni, and I.~Windekilde, ``Business models in
  social networking,'' in \emph{{CMI} {Int}. {Conf}. on {Social} {Networking}
  and {Communities}}, 2009.

\bibitem{BGGS2017}
F.~Beierle, K.~Grunert, S.~G{\"o}nd{\"o}r, and V.~Schl{\"u}ter, ``{T}owards
  {P}sychometrics-based {F}riend {R}ecommendations in {S}ocial {N}etworking
  {S}ervices,'' in \emph{2017 {IEEE} 6th {I}nternational {C}onference on {AI}
  \& {M}obile {S}ervices ({AIMS} 2017)}.\hskip 1em plus 0.5em minus 0.4em\relax
  IEEE, 2017, pp. 105--108.

\bibitem{beierle_towards_2015}
F.~Beierle, S.~G{\"o}nd{\"o}r, and A.~K{\"u}pper, ``Towards a {{Three}}-tiered
  {{Social Graph}} in {{Decentralized Online Social Networks}},'' in
  \emph{Proc. 7th {{International Workshop}} on {{Hot Topics}} in
  {{Planet}}-Scale {{mObile Computing}} and {{Online Social neTworking}}
  ({{HotPOST}})}.\hskip 1em plus 0.5em minus 0.4em\relax {ACM}, Jun. 2015, pp.
  1--6.

\bibitem{smartphoneusage}
A.~Smith, ``{U}.{S}. {S}martphone {U}se in 2015,''
  \url{http://www.pewinternet.org/2015/04/01/us-smartphone-use-in-2015/},
  {A}ccessed 2018-02-15.

\bibitem{farahbakhsh_analysis_2013}
R.~Farahbakhsh, X.~Han, A.~Cuevas, and N.~Crespi, ``Analysis of publicly
  disclosed information in {{Facebook}} profiles,'' in \emph{Proc. 2013
  {{IEEE}}/{{ACM International Conference}} on {{Advances}} in {{Social
  Networks Analysis}} and {{Mining}} ({{ASONAM}})}.\hskip 1em plus 0.5em minus
  0.4em\relax {ACM}, Aug. 2013, pp. 699--705.

\bibitem{beach_whozthat?_2008}
A.~Beach, M.~Gartrell, S.~Akkala, J.~Elston, J.~Kelley, K.~Nishimoto, B.~Ray,
  S.~Razgulin, K.~Sundaresan, B.~Surendar, M.~Terada, and R.~Han,
  ``{{WhozThat}}? {{Evolving}} an {{Ecosystem}} for {{Context}}-{{Aware Mobile
  Social Networks}},'' \emph{IEEE Network}, vol.~22, no.~4, pp. 50--55, 2008.

\bibitem{eagle_social_2005}
N.~Eagle and A.~Pentland, ``Social {{Serendipity}}: {{Mobilizing}} social
  software,'' \emph{Pervasive Computing, IEEE}, vol.~4, no.~2, pp. 28--34,
  2005.

\bibitem{pietilainen_mobiclique:_2009}
A.-K. Pietil{\"a}inen, E.~Oliver, J.~LeBrun, G.~Varghese, and C.~Diot,
  ``{{MobiClique}}: {{Middleware}} for {{Mobile Social Networking}},'' in
  \emph{Proc. {{2nd ACM Workshop}} on {{Online Social Networks}} (WOSN)}.\hskip
  1em plus 0.5em minus 0.4em\relax {ACM}, 2009, pp. 49--54.

\bibitem{yang_e-smalltalker:_2010}
Z.~Yang, B.~Zhang, J.~Dai, A.~Champion, D.~Xuan, and D.~Li,
  ``E-{{SmallTalker}}: {{A Distributed Mobile System}} for {{Social
  Networking}} in {{Physical Proximity}},'' in \emph{2010 {{IEEE}} 30th
  {{International Conference}} on {{Distributed Computing Systems}}
  ({{ICDCS}})}, Jun. 2010, pp. 468--477.

\bibitem{teng_e-shadow:_2014}
J.~Teng, B.~Zhang, X.~Li, X.~Bai, and D.~Xuan, ``E-{{Shadow}}: {{Lubricating
  Social Interaction Using Mobile Phones}},'' \emph{IEEE Transactions on
  Computers}, vol.~63, no.~6, pp. 1422--1433, Jun. 2014.

\bibitem{mei_social-aware_2015}
A.~Mei, G.~Morabito, P.~Santi, and J.~Stefa, ``Social-{{Aware Stateless Routing
  in Pocket Switched Networks}},'' \emph{IEEE Transactions on Parallel and
  Distributed Systems}, vol.~26, no.~1, pp. 252--261, Jan. 2015.

\bibitem{dong_when_2013}
C.~Dong, L.~Chen, and Z.~Wen, ``When {{Private Set Intersection Meets Big
  Data}}: {{An Efficient}} and {{Scalable Protocol}},'' in \emph{Proc. 2013
  {{ACM SIGSAC Conference}} on {{Computer}} \& {{Communications Security}}
  (CCS)}.\hskip 1em plus 0.5em minus 0.4em\relax {ACM}, 2013, pp. 789--800.

\bibitem{kerschbaum_outsourced_2012}
F.~Kerschbaum, ``Outsourced {{Private Set Intersection Using Homomorphic
  Encryption}},'' in \emph{Proc. 7th {{ACM Symposium}} on {{Information}},
  {{Computer}} and {{Comm. Security}} (ASIACCS)}.\hskip 1em plus 0.5em minus
  0.4em\relax {ACM}, 2012, pp. 85--86.

\bibitem{tillmanns_privately_2015}
J.~Tillmanns, ``Privately {{Computing Set}}-{{Union}} and
  {{Set}}-{{Intersection Cardinality}} via {{Bloom Filters}},'' in \emph{20th
  Australasian Conf. on Inf. {{Security}} and {{Privacy}} ({{ACISP}})}, vol.
  9144.\hskip 1em plus 0.5em minus 0.4em\relax {Springer}, 2015, pp. 413--430.

\bibitem{beierle_privacy-aware_2016}
F.~Beierle, K.~Grunert, S.~G{\"o}nd{\"o}r, and A.~K{\"u}pper, ``Privacy-aware
  {{Social Music Playlist Generation}},'' in \emph{Proc. 2016 {{IEEE
  International Conference}} on {{Communications}} ({{ICC}})}.\hskip 1em plus
  0.5em minus 0.4em\relax {IEEE}, May 2016, pp. 5650--5656.

\bibitem{bloom_space/time_1970-1}
B.~H. Bloom, ``Space/{{Time Trade}}-offs in {{Hash Coding}} with {{Allowable
  Errors}},'' \emph{Commun. ACM}, vol.~13, no.~7, pp. 422--426, Jul. 1970.

\bibitem{fan_summary_2000}
L.~Fan, P.~Cao, J.~Almeida, and A.~Z. Broder, ``Summary {{Cache}}: {{A Scalable
  Wide}}-{{Area Web Cache Sharing Protocol}},'' \emph{IEEE/ACM Transactions on
  Networking}, vol.~8, no.~3, pp. 281--293, Jun. 2000.

\bibitem{cormode_improved_2004}
G.~Cormode and S.~Muthukrishnan, ``An {{Improved Data Stream Summary}}: {{The
  Count}}-{{Min Sketch}} and its {{Applications}},'' in \emph{{{LATIN}} 2004:
  {{Theoretical Informatics}}}.\hskip 1em plus 0.5em minus 0.4em\relax
  {Springer}, 2004, pp. 29--38.

\bibitem{jain_using_2005}
N.~Jain, M.~Dahlin, and R.~Tewari, ``Using {{Bloom Filters}} to {{Refine Web
  Search Results}}.'' in \emph{{{WebDB}}}, 2005, pp. 25--30.

\bibitem{schnell_privacy-preserving_2009}
R.~Schnell, T.~Bachteler, and J.~Reiher,
  ``\BIBforeignlanguage{en}{Privacy-preserving record linkage using {{Bloom}}
  filters},'' \emph{\BIBforeignlanguage{en}{BMC Medical Informatics and
  Decision Making}}, vol.~9, no.~1, p.~41, Aug. 2009.

\bibitem{donnet_path_2012}
B.~Donnet, B.~Gueye, and M.~A. Kaafar, ``Path similarity evaluation using
  {{Bloom}} filters,'' \emph{Computer Networks}, vol.~56, no.~2, pp. 858--869,
  2012.

\bibitem{alaggan_blip:_2012}
M.~Alaggan, S.~Gambs, and A.-M. Kermarrec, ``{{BLIP}}: {{Non}}-interactive
  {{Differentially}}-{{Private Similarity Computation}} on {{Bloom}} filters,''
  in \emph{Stabilization, {{Safety}}, and {{Security}} of {{Distributed
  Systems}}}, ser. LNCS, A.~W. Richa and C.~Scheideler, Eds.\hskip 1em plus
  0.5em minus 0.4em\relax {Springer}, 2012, no. 7596, pp. 202--216.

\bibitem{bertin-mahieux_million_2011}
T.~Bertin-Mahieux, D.~P. Ellis, B.~Whitman, and P.~Lamere, ``The million song
  dataset,'' in \emph{Proc. 12th {{International Society}} for {{Music
  Information Retrieval Conference}} ({{ISMIR}})}, 2011.

\end{thebibliography}

\end{document}